\documentclass[10pt,twoside,BCOR7mm,DIV15,headinclude,footexclude,
               cleardoubleempty,idxtotoc]
{scrartcl}

\usepackage{natbib}
\usepackage[font=small,labelfont=bf]{caption}
\usepackage[english]{babel}
\usepackage{graphicx}
\usepackage{hyperref}
\usepackage{scrpage2}
\usepackage{ifthen}
\usepackage{booktabs}
\usepackage{amsmath}
\usepackage{amssymb}
\usepackage{multicol}
\usepackage{float}
\usepackage{hyperref}
\usepackage{breakurl}

\hypersetup{breaklinks=true
,colorlinks=true,linkcolor=black,urlcolor=blue
,citecolor=black}

\addto\captionsenglish{%
}

\makeatletter
\renewcommand{\@biblabel}[1]{}
\renewcommand{\@cite}[2]{%
{#1\ifthenelse{\boolean{@tempswa}}{,#2}{}}}
\makeatother
\ofoot{\thepage}
\ifoot{}

\setcapindent{0em}
\setheadsepline{1pt}

\setkomafont{pagehead}{\normalfont\sffamily}
\setkomafont{pagenumber}{\normalfont\rmfamily}

\makeatletter
\newcommand{\listofcontributions}{\@starttoc{con}}

\newcommand{\l@contribution} {\@dottedtocline{1}{1.5em}{2.3em}}
\makeatother

\newenvironment{contribution}{
\setcounter{section}{0}
\setcounter{figure}{0}
\setcounter{table}{0}
}{
\newpage
\lehead{}
\rohead{}
}

\begin{document}

\setlength{\baselineskip}{2.5ex}

\begin{contribution}
% EXAMPLE AND TEMPLATE FILE FOR PROCEEDINGS OF THE WOLF-RAYET WORKSHOP.
% PLEASE REPLACE THE TEMPLATE TEXT BY YOUR OWN ARTICLE.
% NOTE THAT YOU MUST NOT PROCESS THIS FILE, BUT THE MASTER FILE:
% latex masterfile; dvips masterfile

% RUNNING AUTHOR: PUT AUTHOR NAMED HERE
\lehead{T.~I.\ Madura et al.}

% RUNNING TITLE; SHORTEN THE TITLE IF NECESSARY
% IN CASE OF A ONE-PAGE CONTRIBUTION (POSTER),
% SQUEEZE AUTHORS AND TITLE IN THIS LINE (Author: Title ...)
\rohead{3D modeling $\eta$~Car's colliding winds}

\begin{center}
% FULL TITLE HEADING
{\LARGE \bf 3D hydrodynamical and radiative transfer modeling of $\eta$~Carinae's colliding winds}\\
\medskip

% AUTHORS LIST
{\it\bf T.~I. Madura$^1$, N.\ Clementel$^2$,  T.~R.\ Gull$^3$, C.~J.~H.\ Kruip$^4$, J.-P. Paardekooper$^5$ \& V.\ Icke$^4$}\\

% AFFILIATIONS
{\it $^1$USRA \& NASA Goddard Space Flight Center, Greenbelt, MD, USA}\\
{\it $^2$South African Astronomical Observatory, Cape Town, South Africa}\\
{\it $^3$NASA Goddard Space Flight Center, Greenbelt, MD, USA}\\
{\it $^4$Leiden Observatory, Netherlands}\\
{\it $^5$Universit\"at Heidelberg, Germany}\\

% ABSTRACT
\begin{abstract}
We present results of full 3D hydrodynamical and radiative transfer simulations of the colliding stellar winds in the massive binary system $\eta$~Carinae. We accomplish this by applying the SimpleX algorithm for 3D radiative transfer on an unstructured Voronoi-Delaunay grid to recent 3D smoothed particle hydrodynamics (SPH) simulations of the binary colliding winds. We use SimpleX to obtain detailed ionization fractions of hydrogen and helium, in 3D, at the resolution of the original SPH simulations. We investigate several computational domain sizes and Luminous Blue Variable primary star mass-loss rates. We furthermore present new methods of visualizing and interacting with output from complex 3D numerical simulations, including 3D interactive graphics and 3D printing. While we initially focus on $\eta$~Car, the methods employed can be applied to numerous other colliding wind (WR~140, WR~137, WR~19) and dusty `pinwheel' (WR~104, WR~98a) binary systems. Coupled with 3D hydrodynamical simulations, SimpleX simulations have the potential to help determine the regions where various observed time-variable emission and absorption lines form in these unique objects.
\end{abstract}
\end{center}

% TEXT OF THE PAPER, TWO-COLUMN STYLE
\begin{multicols}{2}

\section{3D SPH Simulations of $\eta$~Car's Colliding Winds}

As already discussed in these proceedings (e.g. Gull, Morris et al., Corcoran et al., Hamaguchi et al.), Eta Carinae ($\eta$~Car) and its surrounding bipolar Homunculus nebula constitute an ideal astrophysical laboratory for the study of massive binary interactions and evolution, and stellar wind-wind collisions. Recent three-dimensional (3D) simulations set the stage for understanding the complex time-dependent 3D stellar wind outflows in $\eta$~Car. Spurned by recent suggestions that the mass-loss rate of the Luminous Blue Variable (LBV) primary star might have dropped by a factor of $2 - 3$ over roughly the past decade \citep{corcoran10,mehner12}, \cite{madura13} presented results from a series of 3D smoothed particle hydrodynamics (SPH) simulations of $\eta$~Car's binary colliding winds assuming three different primary star mass-loss rates and using various computational domain sizes.

These SPH simulations reveal that the value of the primary's mass-loss rate greatly affects the time-dependent hydrodynamics of the wind-wind collision at all spatial scales investigated ($\sim$15.5~AU to 1550~AU). The SPH simulations also show that the post-shock wind of the companion star switches from the adiabatic to the radiative-cooling regime during periastron passage ($\phi \approx 0.985 - 1.02$). This switchover starts later and ends earlier the lower the value of the primary's mass-loss rate and is caused by the encroachment of the primary wind into the acceleration zone of the companion's wind, plus radiative inhibition of the companion's wind by the more luminous primary. This cooling switchover has important implications for understanding the peculiar behavior of $\eta$~Car's X-ray light curve during the past two periastron passages, which showed an inconsistent recovery from the X-ray minimum \citep[][Corcoran et al., these proceedings]{corcoran10}.

Understanding the behavior of the post-shock companion wind in $\eta$~Car is crucial for understanding the behavior of the X-ray light curve, since it is in the extremely hot (up to $\sim$10$^{8}$~K) post-shock secondary wind that the majority of the hard X-rays observed are generated. As discussed in detail in \cite{madura13}, the varying duration of $\eta$~Car's X-ray minimum from cycle-to-cycle likely depends strongly on the length of time the post-shock secondary wind can remain in the radiative cooling regime during periastron passage. For most of $\eta$~Car's highly eccentric orbit ($e \sim$~0.9), the post-shock companion wind cools adiabatically and thus remains relatively hot, generating hard X-rays. However, the cooling parameter $\chi$ that determines whether the post-shock gas is in the adiabatic or radiative cooling regime is proportional to the pre-shock wind speed to the fourth power \citep{stevens92}. Thus, even a small change in the pre-shock wind speed of the companion star can have a dramatic effect on the cooling of the post-shock gas.

Normally, we think of the radiatively-driven wind from a massive star as following a so-called $\beta$-law, wherein the wind speed at radius $r$ is given by $v(r) = v_{\infty}(1 - R_{\star}/r)^{\beta}$, where $v_{\infty}$ is the wind terminal speed and $R_{\star}$ the stellar radius. However, as discussed in \cite{madura13}, because of the highly elliptical orbit and presence of the incredibly luminous primary star, the pre-shock companion wind speed in the $\eta$~Car system is a complicated function of the stellar luminosity ratio and stellar separation. The pre-shock wind speed of the companion star in $\eta$~Car decreases by roughly a factor of three between apastron (phase 0.5) and phase 0.99 \citep{madura13}. Because the cooling parameter $\chi$ depends on the wind speed to the fourth power, a factor of three reduction in the companion's wind speed around periastron causes the post-shock gas to switch from being adiabatic to strongly radiative. This reduces greatly the volume of hot X-ray generating gas, and thus the observed X-ray flux. Only after periastron when the stellar separation starts to increase does the pre-shock companion wind speed also increase, and the post-shock gas can switch back to the adiabatic regime, again producing hard X-rays. The duration of the radiative cooling phase of the post-shock companion wind thus very likely affects the duration of the observed X-ray minimum and timing of the X-ray recovery following periastron.

Since the primary star in $\eta$~Car is a known LBV subject to periods of instability, and because the wind-wind collision region is very chaotic and dynamic due to the various instabilities that arise at the interface between the two colliding winds \citep{stevens92,madura13}, it is reasonable to suggest that such instabilities and/or small variations in stellar/wind parameters can lead to changes in the overall pre-shock wind speeds, location of wind momentum balance, and duration of strong radiative cooling in the post-shock companion wind. Thus, changes in the duration of the X-ray minimum and timing of the X-ray recovery following periastron may be natural in a system like $\eta$~Car.

\section{3D Radiative Transfer Models}

In addition to the `current' colliding winds interaction responsible for the X-ray emission, which occurs in the innermost regions of the $\eta$~Car system (at spatial scales comparable to the orbital semimajor axis length $a = 15.4$~AU), larger scale ($\sim 3250$~AU diameter) 3D SPH simulations reveal outer wind-wind collision regions (WWCRs) that extend thousands of AU from the central stars \citep{madura12,madura13}. Long-slit spectral observations with the Hubble Space Telescope/Space Telescope Imaging Spectrograph (HST/STIS) reveal these spatially-extended WWCRs, seen via emission from multiple low- and high-ionization forbidden lines \citep{gull09,gull11,teodoro13}. Observations of the different broad high- and low-ionization forbidden emission lines provide an excellent tool to constrain the orientation of the system, the primary's mass-loss rate, and the ionizing flux of the hot secondary star. In addition to emission from forbidden lines, helium presents several interesting time-variable emission and absorption features that provide important clues on the geometry and physical proprieties of the system and the individual stars.

To set the stage for future efforts aimed at generating synthetic observations for comparison with available and future HST/STIS and ground-based data, we have performed a series of 3D radiative transfer simulations of the interacting winds in $\eta$~Car. We use the SimpleX algorithm to post-process the 3D SPH simulation output and obtain detailed 3D maps of the ionization fractions of hydrogen and helium. The resultant ionization maps constrain the regions where observed forbidden emission lines can form, and the regions where helium is singly- and doubly-ionized.

Our SimpleX results recently appeared in the series of papers \cite{clementel14, clementel15a, clementel15b}. In \cite{clementel14}, we focus on large-domain ($\sim$3250~AU diameter) simulations at an orbital phase of apastron, investigating the 3D ionization structure of the extended colliding wind structures assuming different primary mass-loss rates. These simulations are comparable in size to the HST/STIS observations of the extended forbidden emission structures \citep{gull09,gull11,teodoro13}. We find that at apastron, the dense primary wind confines the photoionization, and hence the forbidden line emission, to the secondary star's (and observer's) side of the system. Changing the primary's mass-loss rate results in quite different ionization volumes, with much more ionized gas present for lower mass-loss rates. The large apparent changes in ionization volume with decreasing primary mass-loss rate imply that any major decrease in primary mass-loss should lead to significant observable changes in the spatial extent, location, and flux (proportional to density squared) of the broad high-ionization forbidden emission lines. Future models based on the SimpleX results may be used to constrain any such potential mass-loss changes.

\cite{clementel15a} and \cite{clementel15b} investigate the ionization structure of helium in $\eta$~Car's innermost regions ($\lesssim$310~AU in diameter) at apastron and periastron, respectively, for different primary mass-loss rates. We find that reducing the primary's mass-loss rate increases the volume of He$^{+}$. Doubly-ionized helium (He$^{++}$) only exists in the extremely hot post-shock secondary wind. At orbital phases near apastron, lowering the primary mass-loss rate produces large variations in the volume of He$^{+}$ in the pre-shock primary wind located on the periastron side of the system. Our results show that binary orientations in which the secondary star at apastron is on our side of the system are most consistent with the available observations.

At an orbital phase of periastron, our simulations show that generally, He$^{0+}$-ionizing photons from the companion star are not able to penetrate into the pre-shock primary wind, and that He$^{+}$ due to the companion's ionizing flux is only present in a thin layer along the leading arm of the inner WWCR and in a small region close to the stars. Lowering the primary's mass-loss rate allows the companion's ionizing photons at periastron to reach the expanding unshocked secondary wind on the apastron side of the system, and create a low fraction of He$^{+}$ in the pre-shock primary wind. With the companion at apastron on our side of the system, our results are qualitatively consistent with the observed variations in strength and radial velocity of $\eta$~Car's helium emission and absorption lines.

\section{3D Visualization and Printing}

3D hydrodynamical and radiative transfer simulations of $\eta$~Car show that the complex time-varying WWCR has a major impact on the observed X-ray emission, the optical and ultraviolet light curves and spectra, and the interpretation of various line profiles and interferometric observables \citep[see][and references therein]{madura15}. While 3D simulations have helped to increase substantially our understanding of the present-day $\eta$~Car binary, historically we have been limited by an inability to adequately visualize the full 3D time-dependent geometry of the WWCR. However, this is crucial if we are to thoroughly understand how and where various forms of observed emission and absorption originate.

In an effort to better understand the 3D structure of $\eta$~Car's time-dependent WWCRs, \cite{madura15} recently published 3D interactive visualizations and 3D prints of $\eta$~Car's colliding stellar winds at orbital phases of apastron, periastron, and three months after periastron. The figures in the PDF version of \cite{madura15}, downloadable directly from the NASA ADS service (\href{http://adsabs.harvard.edu/abs/2015MNRAS.449.3780M}{\texttt{http://adsabs.harvard.edu/}}), are fully 3D interactive when using the standard Adobe Reader (\href{https://get.adobe.com/reader/}{\texttt{https://get.adobe.com/reader/}}). The 3D models presented in \cite{madura15} also constitute the first 3D printable output from a supercomputer simulation of a complex astrophysical system.

The 3D visualizations and prints in \cite{madura15} reveal important, previously unknown `finger-like' structures at orbital phases shortly after periastron ($\sim 1.045$) that protrude radially outwards from the spiral WWCR. We speculate that these fingers are related to instabilities (e.g. thin-shell, Rayleigh�Taylor) that arise at the interface between the radiatively cooled layer of dense post-shock primary-star wind and the faster ($\sim 3000$~km/s), adiabatic post-shock companion-star wind. The 3D visualizations also reveal that the inability of the companion star's wind to collide with the prmiary's downstream wind produces a large `hole' in the trailing arm near the WWCR apex at periastron. The size of this hole is directly connected with the wind momentum ratio, and therefore the WWCR opening angle. As expected, higher primary mass-loss rates cause a larger hole than lower primary mass-loss rates. After periastron, this hole expands and propagates downstream along the vanishing WWCR trailing arm.

The success of the work in \cite{madura15} and the relatively easy identification of previously unrecognized physical features highlight the important role 3D printing and interactive 3D graphics can play in the visualization and understanding of complex 3D time-dependent numerical simulations of astrophysical phenomena. The 3D printable models of $\eta$~Car's colliding winds and the $\eta$~Car Homunculus nebula \citep{steffen14} are freely available online at the NASA 3D Resources website \href{http://nasa3d.arc.nasa.gov/}{\texttt{http://nasa3d.arc.nasa.gov/}}.

\bibliographystyle{aa} % style aa.bst
%\bibliography{myarticle}

\end{multicols}

\end{contribution}

%%-------------------------------------------------------

\end{document}